\documentclass[final,3p,times,twocolumn]{elsarticle}
\usepackage{lineno,hyperref}
\usepackage{color}
\modulolinenumbers[5]

\journal{Journal of \LaTeX\ Templates}

\begin{document}

\begin{frontmatter}

\title{Atomic resolution mapping of phonon excitations in STEM-EELS experiments}



\author[mysecondaryaddress]{R. Egoavil}
\cortext[mycorrespondingauthor]{Corresponding author}
\ead{ricardo.egoavil@uantwerpen.ua.ac.be}
\author[mysecondaryaddress]{N. Gauquelin}
\author[mysecondaryaddress]{G. T. Martinez}
\author[mysecondaryaddress]{S. Van Aert}
\author[mysecondaryaddress]{G. Van Tendeloo}
\author[mysecondaryaddress]{J. Verbeeck}


\address[mysecondaryaddress]{EMAT, University of Antwerp, Department of Physics\\
Groenenborgerlaan 171, B-2020 Antwerp, Belgium}

\begin{abstract}

Atomically resolved electron energy-loss spectroscopy experiments are commonplace in modern aberration-corrected transmission electron microscopes. Energy resolution has also been increasing steadily with the continuous improvement of electron monochromators. Electronic excitations however are known to be delocalised due to the long range interaction of the charged accelerated electrons with the electrons in a sample. This has made several scientists question the value of combined high spatial and energy resolution for mapping interband transitions and possibly phonon excitation in crystals. In this paper we demonstrate experimentally that atomic resolution information is indeed available at very low energy losses around 100 meV expressed as a modulation of the broadening of the zero loss peak. Careful data analysis allows us to get a glimpse of what are likely phonon excitations with both an energy loss and gain part. These experiments confirm recent theoretical predictions on the strong localisation of phonon excitations as opposed to electronic excitations and show that a combination of atomic resolution and recent developments in increased energy resolution will offer great benefit for mapping phonon modes in real space.
\end{abstract}

\begin{keyword}
\texttt{EELS, Phonons, Monochromator}

\end{keyword}

\end{frontmatter}

\linenumbers

\section{Introduction}
Recent developments in aberration-corrected transmission electron microscopy as well as improvements in spectrometers and monochromators \cite{Tiemeijer_2012} provide us with instruments that are capable of obtaining atomic resolution combined with electron energy loss spectroscopy (EELS) at energy resolutions of the order of 100~meV  \cite{Tiemeijer_1999, Freitag_2005,H_A_Brink_2003}, and recently even approaching 10~meV \cite{Krivanek_2008}. Typical experiments focus on surface plasmons with modes as low as 0.17~eV being experimentally accesible \cite{Rossouw_2013}. It is also possible to probe changes in the band-gap of semiconductors at the atomic scale in semiconductor devices \cite{Benthem_2006,Benthem_2001}. Improving the energy resolution further seems attractive in order to study phonon lattice vibrations which typically occur between a few meV and 1~eV at large scattering angles (10-1000~mrad). So far, this signal has not been (yet) resolved in standard transmission electron microscopy. Phonon spectra are however routinely studied with electron energy loss spectra on dedicated surface sensitive HREELS instruments \cite{A_D_Baden_1981}.
Combining this capability with atomic resolution seems attractive as it would allow to locally observe phonon modes which is especially attractive at interfaces and defects.
The common thinking in the EELS community so far has been that low loss EELS and spatial resolution contradict each other due to the effect of delocalisation. Delocalisation allows an electronic excitation to be excited even though the fast electron is some distance away from the scattering center due to long range coulomb interaction between the fast electron and the electrons making up the scatterer. This delocalisation is described and verified in great detail and was found to scale approximately inversely proportional to the energy loss \cite{Muller&Silcox_1995,Egerton_2003,Verbeeck_2005}. From this point of view, spatial resolution would suffer dramatically when going from an energy loss of hundreds of eV's (core loss EELS, typical delocalisation smaller than the interatomic distance) to losses below 1~eV. Typically this delocalisation argument holds in free space at edges of the sample, but its validity is strongly reduced inside materials where screening can strongly reduce its effect.
It is important to keep in mind that this type of delocalisation is derived for electronic transition where the fast electron excites the sample electrons to higher lying states. For phonon excitations, which are of major interest in the region below 1~eV, the situation is quite different. Here, the fast electron couples to a lattice vibration mode via Coulomb interaction, but now the delocalisation is strongly reduced due to the extreme difference in mass between the fast electron and the lattice atoms. This results in high angle scattering (10-100~mrad) and strong localisation of the scattering.

Generalized phonon density of states in cubic $\mathrm{SrTiO_{3}}$ were computed,  the phonon spectral range extends from 0-120~meV, where the spectral signatures at the end of this energy range are due to the Sr-O and Ti-O bond vibrations \cite{Narayani_Choudhury_2008}. In addition, surface optical phonons on $\mathrm{SrTiO_{3}}$ were detected below 0.1~eV (57 and 92~meV) by high-resolution EELS (HREELS), and multiphonon excitations are also observed around 0.2-0.3~eV \cite{A_D_Baden_1981}. Measurements were performed by conventional techniques like infrared spectroscopy, ultraviolet Raman spectroscopy and HREELS, which are bulk and surface measurements \cite{ A_D_Baden_1981, D.A.Tenne_2010, Sarbajit_Banerjee_2006}. However, the spatial resolution is a limiting factor of these techniques, with an expected resolution of a few nanometers at best \cite{Egerton_2003}. More recently, localization of vibrational excitations by a high-energy electron beam have been shown to be theoretically possible by Dwyer \cite{Christian_Dwyer_2014}. The vibrational EELS images of $\mathrm{H_{2}}$ and $\mathrm{CO}$ molecules were computed with atomic-scale spatial resolution, by using the "so-called" M$\ddot{o}$ller potential for excitation of the vibrational modes.

On the other hand, preservation of elastic contrast in low-loss EELS mapping has been reported by S. Lazar et al. \cite{lazar_imaging_2010}, where the filtered image of the zero loss peak (ZLP) intensity shows the complementary nature of the high angle annular dark field (HAADF) intensity and the elastic contrast. Furthermore, atomically resolved signatures were observed at 3~eV at high collection angle (124~mrad) and addressed as being possibly related to phonon assisted losses. In this paper, we will demonstrate that indeed subtle changes in the low loss region of an EELS spectrum recorded on a canonical sample of $\mathrm{SrTiO_{3}}$ occur when scanning a focussed electron probe of atomic size on the different atomic columns. We center our attention on changes in the low-energy loss region, most remarkably below 0.5~eV. We argue the possibility of mapping "optical multiphonon states" on $\mathrm{SrTiO_{3}}$ whitin the energy range between 0.14-0.5~eV in agreement with literature \cite{A_D_Baden_1981,Narayani_Choudhury_2008}. This work should encourage the further development of very high resolution spectrometers and monochromators and shows that the combination with atomic resolution is very relevant for the study of phonon behaviour near imperfections and interfaces.

\section{Experimental}
 A monocrystalline substrate of $\mathrm{SrTiO_{3}}$ with $\sim$18-20~nm thickness was investigated using a TEM lamella prepared perpendicular to the [100] zone axis orientation by focused ion beam milling. HAADF imaging, Scanning transmission electron microscopy (STEM) and EELS experiments were performed using the QuAntEM microscope at the University of Antwerp. This is an FEI Titan$^3$ microscope, equipped with an aberration corrector for both image and probe forming lenses, and a monochromator to optimize the energy resolution for EELS measurements up to 120~meV, as determined from the full width at half maximum (FWHM) of the ZLP. As we will be looking at very subtle changes near the zero loss peak, we performed experiments  at different collection angles to study the effect of possible unwanted spectrometer aberrations. Cerenkov radiation effects were limited by working with thin samples at an accelaration voltage of 120~keV where spatial resolution performance is approx. $\sim$1.3(4)~$\AA$ (probe size) at a convergence semi-angle of 21~mrad. \cite{Erni_2008,Kulpreet_2013}. Experiments performed at 300~kV under similar conditions show strong agreement with these observations (see supplementary information). The collection semi-angle $\beta$ of the spectrometer was set to $\sim$38, 43, 129 and 225~mrad, respectively.
ZLP extraction at the tail of the ZLP is notoriously difficult and error prone. To overcome this, we abandoned direct background subtraction or the use of pre-recorded ZLP spectra in favor of two alternative methods:
\begin{itemize}
\item{Each spectrum in a spectrum image is first energy drift corrected, scaled to its maximum and then \emph{divided} by the average spectrum obtained by summing all spectra, also scaled to the maximum. This method is extremely sensitive to changes in the spectral shape of the ZLP but leads to spectra that are somewhat hard to interpret as they show the fractional deviation from the average spectrum. Also noise amplification starts to become important for energies where the average spectrum is low.}
\item{Each spectrum is energy drift corrected, scaled to its maximum and then a scaled version of the average spectrum is \emph{subtracted}. Which the scaling chosen such that the difference is zero on the maximum of the ZLP. This leads to more directly interpretable spectra showing the increase or decrease of certain features near the ZLP depending on the spatial position of the probe.}
\end{itemize}

\section{Results and Discussion}
\begin{figure}[ht]
    \centering
        \includegraphics[width=1.0\linewidth]{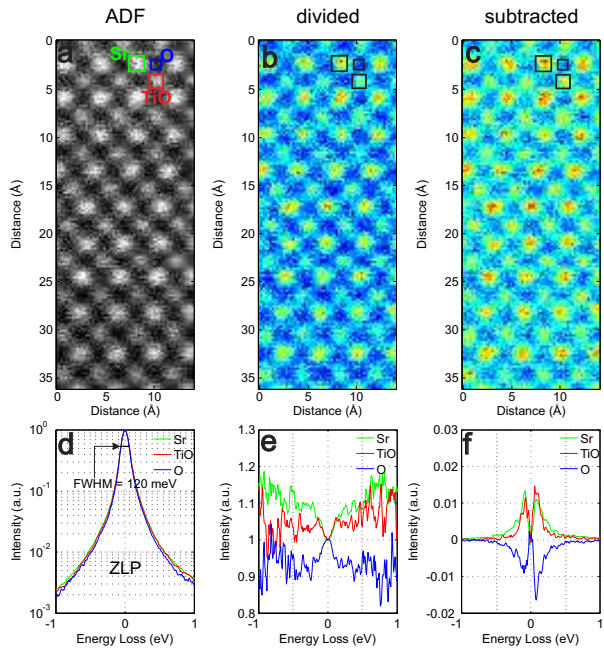}
    \caption{2D EELS maps acquired at 120~kV and $\beta$ = 129~mrad on a $\mathrm{SrTiO_{3}}$ lamella. (a) Recorded HAADF image and (b,c) its corresponding \emph{divided} and \emph{subtracted} EELS maps. (d) logarithmic scaled zero loss spectra and (e,f) spectra line profiles on 3 different regions for Sr (green), TiO (red) and O (blue). The FWHM on the zero loss spectra indicates the corresponding energy resolution of experiment (120~meV). The b and c maps are both integrated over an energy range of 0.14 to 0.5~eV.}
    \label{fig1}
\end{figure}

Using STEM EELS, 2D spectrum images were acquired at different acceleration voltage and collection angles. A resulting HAADF image for experiments performed at 120~keV and $\beta$ = 129~mrad, and its corresponding \emph{divided} and \emph{subtracted} 2D maps treated as described previously are shown in Figure~\ref{fig1}(a-c). The following panels (d-f) present normalised spectra taken from the 3 different positions indicated on  Figure~\ref{fig1}b, corresponding to Sr (green), TiO (red) and O (blue) atomic columns, respectively. One clearly recognises a subtle broadening and shrinking of the width of the ZLP when comparing the 3 different spectra shown on a logarithmic scale of the electron intensity (Figure~\ref{fig1}d). Figures~\ref{fig1}(e,f) show the \emph{divided} and \emph{subtracted} spectra profiles in the range of energy  -1 to 1~eV. Some deviation from the averaged spectrum extending to 0.5~eV on each side of the ZLP can be noticed.

\begin{figure}[ht]
    \centering
        \includegraphics[width=1.0\linewidth]{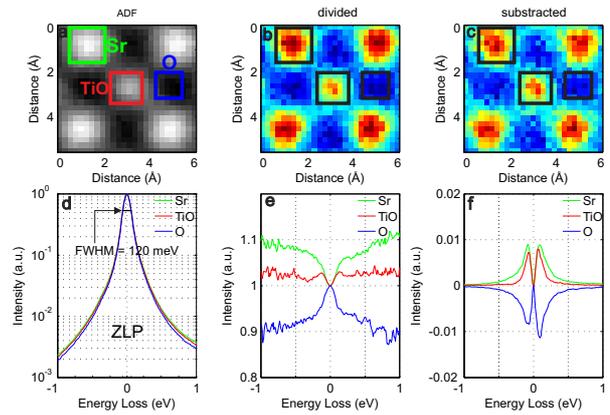}
    \caption{Averaged unit cell showing the (c,d) \emph{divided} and \emph{subtracted} EELS maps and (e,f) its corresponding integrated spectra profiles after averaging over 16 unit cells.}
    \label{fig2}
\end{figure}

In order to better filter out the changes in the spectral shape of the ZLP and improve the signal to noise ratio of the experiment, unit cell averaging was applied, averaging over a total of 16 unit cells. The resulting average unit cell HAADF image is shown in Figure~\ref{fig2}a, together with the spectral shape of the ZLP for a region belonging to a Sr (green), TiO (red) and O (blue) column, extracted in a similar manner as for Figure~\ref{fig1}. From these profiles we observe that the ZLP shows symmetric signatures in both sides of the elastic peak, \emph{gain} and \emph{loss} energy regions. A strong positive deviation from the average spectrum on Sr columns up to almost 800 meV is observed on each side of the ZLP, whereas a less intense positive deviation is observed on TiO columns extending only 300~meV on both sides. On pure oxygen columns a negative deviation similar to the ones observed on the Sr columns can be noticed for both data processing. This averaging improves the signal to noise ratio (SNR) in order to better observe the subtle changes when moving the electron probe over the different positions within a unit cell. The results are summarized in Figure~\ref{fig2}.

\begin{figure}[ht]
    \centering
        \includegraphics[width=1.0\linewidth]{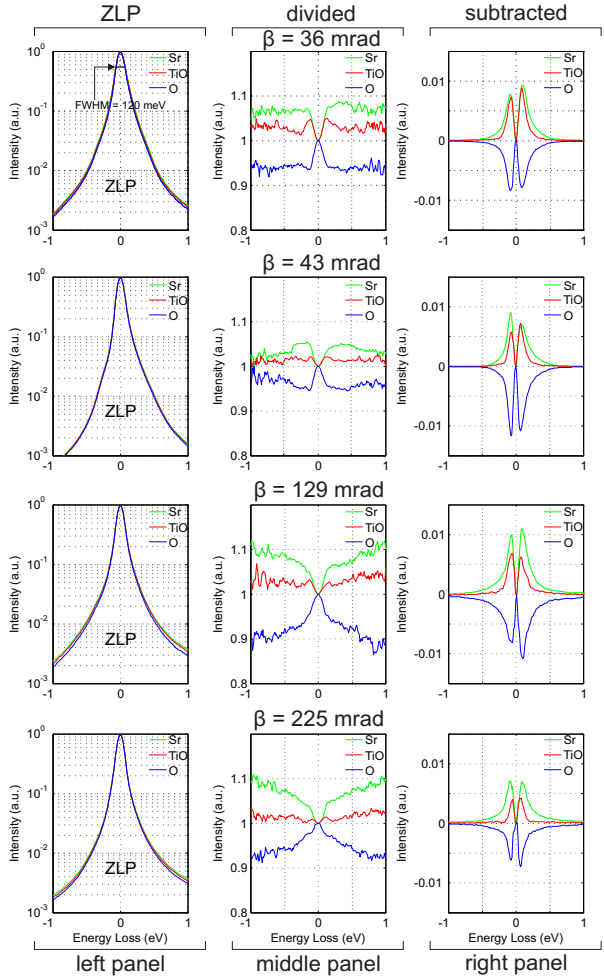}
    \caption{(left to right)The zero loss spectra and its corresponding average spectrum \emph{divided} and \emph{subtracted} EELS spectrum profiles at (top to bottom) different collection semi-angles $\beta$ = 36~mrad, $\beta$ = 43~mrad, $\beta$ = 129~mrad and $\beta$ = 225~mrad. Results are presented after averaging over 22, 44, 16 and 95 unit cells of each dataset, respectively.}
    \label{fig3}
\end{figure}

The same integration was applied on spectral datasets recorded at different collection semi-angles and the results are presented in Figure~\ref{fig3}. The experiments show fundamentally the same effects demonstrating a reproducible but subtle change in the shape of ZLP with atomic resolution. The broadening of the ZLP is observed to be highest on a Sr column, intermediate on a Ti-O column and lowest on an oxygen column on Figure~\ref{fig3}(left panels). We argue that changes in the energy-gain region in the middle and right panels of Figure~\ref{fig3}, can be attributed to thermally excited phonons as recently discussed by \cite{García_de_Abajo_2008, Asenjo_Garcia_2013} and the signatures between 0.14-0.5~eV might be strongly related to phonon mediated energy losses, confirming the possibility to observe multiple phonon states in this energy range  as reported by Baden et.al \cite{A_D_Baden_1981}. Note that for a small collection semi-angle $\beta$ = 36~mrad, the intensity of the Sr and TiO signatures present similar magnitude in qualitative agreement with calculations done by Dwyer \cite{Christian_Dwyer_2014} on a $\mathrm{CO}$ molecule.

\begin{figure}[ht]
    \centering
        \includegraphics[width=1.0\linewidth]{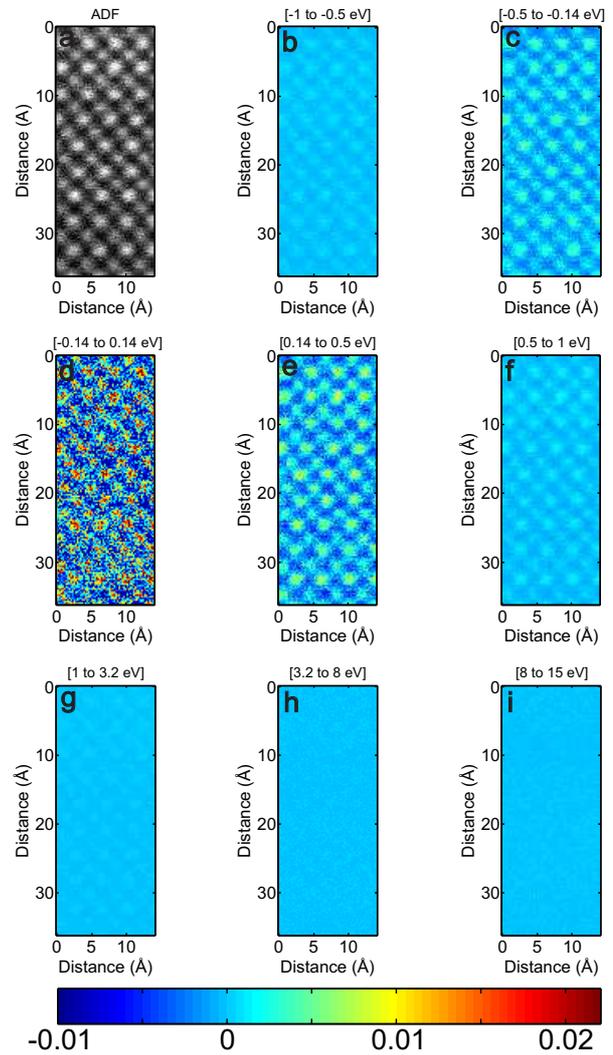}
    \caption{HAADF image and its corresponding selected \emph{subtracted} spectrum-image integrated over different energy ranges of [-1.0 to -0.5~eV]; [-0.5~eV to -0.14~eV]; [-0.14 to 0.14~eV]; [0.14 to 0.5~eV]; [1.0 to 3.2~eV]; [3.2 to 8.0~eV] and [8.0 to 15~eV]. Images are plotted using the same color scale.}
    \label{fig4}
\end{figure}

In order to understand more about the significance of this atomic scale variation of the ZLP, one can integrate the resulting \emph{subtracted} spectrum-image in the following energy ranges [-1.0 to -0.5~eV]; [-0.5~eV to -0.14~eV]; [-0.14 to 0.14~eV]; [0.14 to 0.5~eV]; [1.0 to 3.2~eV]; [3.2 to 8.0~eV] and [8.0 to 15~eV] as presented on Figure~\ref{fig4}(b-i). Figure~\ref{fig4}g shows attenuation of the contrast observed close to the ZLP in the range of [1.0 to 3.2~eV], below the bandgap region for $\mathrm{SrTiO_{3}}$. Figures~\ref{fig4}(h,i) show the absence of atomic contrast on the spectral range corresponding to the interband and plasmon energy losses respectively in good agreement with the expected delocalisation limits for electronic excitations \cite{Benthem_2006, lazar_imaging_2010}.

We conclude that low loss EELS spectra demonstrate a clear atomic resolution signal at very low energy losses which are usually hidden in the tails of the ZLP. We argue that this spectral feature is most likely related to phonon excitations for the following reasons:

\begin{itemize}
\item{The energy range agrees with multiple optical phonon excitations as observed with other techniques in $\mathrm{SrTiO_{3}}$ \cite{A_D_Baden_1981,Narayani_Choudhury_2008}.}
\item{Interband transitions in this material are expected at higher energies between 1 to 15~eV) \cite{Benthem_um_2001, Benthem_2006}}
\item{Typical plasmon excitations are also expected in a higher energy range form  4-50~eV \cite{Yamazaki_2010}.}
\item{The fact that these fluctuations have atomic localisation strongly indicates that they are massive lattice vibrations rather than the more common electronic excitations.}
\item{An energy \emph{gain} spectrum is clearly present which can be interpreted as coming from thermally populated phonon modes interacting with the fast electron (experiment at room temperature).\cite{A_D_Baden_1981}}
\item{The atomic resolution contrast only appears near the zero loss peak but is not observed for the region of interband transitions or plasmons where delocalisation makes the spectrum independent on where the atom probe hits.}
\end{itemize}

Note that the above observations seem to contradict with the presence of an inverted contrast that was labeled as elastic contrast by Lazar et al. \cite{lazar_imaging_2010}. However this is only an apparent contradiction as we chose to highlight spatial changes in inelastic scattering probability by normalizing our spectra to the maximum in the ZLP. This procedure effectively cancels all elastic scattering contribution and leaves only those contributions that change the shape of the EELS spectrum while being insensitive to a pure scaling. For comparison we show in supplementary information that the elastic contrast is also present in our case when we normalize to the total electron counts instead of the maximum in the ZLP for each spectrum, in agreement with Lazar et al.
As the observed changes in spectral shape are subtle and as we can not resolve true single phonon excitation modes with the current energy resolution of our instrument, we will also discuss possible sources of artefacts that could lead to a similar result:

\begin{itemize}
\item{\emph{Nonlinearity of the CCD detector:} Our experiments are well outside the range of saturation with the highest signal in all spectra remaining under 50$\%$ of the dynamic range of the camera ($<$35000 counts). Moreover, the spatially dependent spectral differences are in general assymetric with respect to the zero loss peak, depending on the collection angle as observed in fig.\ref{fig3}. This can not be caused by only a nonlinear response damping the maximum of the ZLP.}
\item{\emph{Afterglow of the scintilator:} The results are found to be essentially the same for a wide range of exposure times and illumination intensities. A variation of the exposure time and illumination intensities from 1 to 15~ms and 40 to 80~pA did not show any variation of the observed signatures excluding afterglow as an important artefact in our data. Moreover, afterglow would affect the spectra taken along the scan direction in a different than those perpendicular to the scan direction. We observe no such anisotropy in our data.}
\item{\emph{Spectrometer aberration mediated by a different scattering angle profile for heavy columns compared to light columns:} Measurements obtained at different collection angles ranging from 36 to 225~mrad did not show a significant variation of the observed atomically resolved spectral signatures. Also changing the entrance aperture from 5 to 2.5~mm does not affect the recorded signal significantly. Note that all our experiments are in a range where the convergence angle is considerably smaller than the collection angle and the considerably larger than the effective scattering angle $\theta$E (a few tens of mrads at 1~eV typically)\cite{Egerton_2011}. This brings our experiments in the range where Dwyer estimates the phonon contrast to appear on the atom columns \cite{Christian_Dwyer_2014}.}
\end{itemize}

The results presented here have to be contrasted to atomic resolution that occurs due to preservation of elastic contrast as demonstrated e.g. in \cite{lazar_imaging_2010}. In this case, the inelastic excitation is heavily delocalised creating a coherent energy loss wave that can interact elastically with the rest of the crystal thereby gaining atomic resolution similar to TEM images. Such preservation of elastic contrast is a purely coherent effect and would be dominant for small collection angles where STEM-EELS approaches EFTEM \cite{Verbeeck_2009}. Our experiments show better contrast for increasing collection angles, indicating an incoherent scattering process.

The above discussion makes it likely that we are indeed observing phonon excitation with atomic scale localisation, in agreement with recent theoretical simulations by Dwyer \cite{Christian_Dwyer_2014} and in agreement with the fact that localisation is expected to be much higher than for electronic excitations.

In the particular case of the present study, outlining (multi)phonon states on $\mathrm{SrTiO_{3}}$ below 0.5~eV in rather good agreement with previously reported measurements \cite{A_D_Baden_1981}.

\section{Conclusions}
In this work we have shown that inelastic excitations with energy losses in the range of 100~meV still provide spatial resolution at the atomic scale. We argue that the excitations we see are phonon excitations which are far more localised as compared to electronic transitions due to the mass difference between the fast electrons and the lattice atoms. We observe both loss and gain contributions to the signal but energy resolution was not sufficient to pinpoint individual phonon loss modes. These observations demonstrate however that indeed improved monochromators combined with aberration corrected STEM instruments will lead to local probing of phonon states with delocalisation not being a limiting factor.

\section{Acknowledgments}
This work is supported by the European Union Council under the 7th Framework Program (FP7) grant number 246102 IFOX, which are gratefully acknowledged. J.V. acknowledges funding from the European Research Council under the 7th Framework Program (FP7), ERC starting grant number 278510 VORTEX. J.V. and G.V.T acknowledge funding from the European Research Council under the 7th Framework Program (FP7), ERC grant number 246791-COUNTATOMS. The titan microscope was partly funded by the Hercules fund from the Flemish Government. The authors acknowledge financial support from the European Union under the Seventh Framework Program under a contract for an Integrated Infrastructure Initiative. Reference No. 312483-ESTEEM2.

\section*{References}

\section{Supplementary information}

\bibliography{mybibfile}

\end{document}